\documentclass[aps,twocolumn,showpacs]{revtex4}
\usepackage{amsmath}
\usepackage{graphicx}

\begin{document}

\title{Classical Hamiltonian Dynamics and Lie Group Algebras}
\author{B. Aycock, A. Roe, J. L. Silverberg and A. Widom}
\affiliation{Physics Department, Northeastern University, Boston MA 02115}

\begin{abstract}
The classical Hamilton equations of motion yield a structure 
sufficiently general to handle an almost arbitrary set of ordinary differential 
equations. Employing elementary algebraic methods, it is possible within 
the Hamiltonian structure to describe 
many physical systems exhibiting Lie group symmetries. Elementary examples 
include magnetic moment precession and the mechanical orbits of color 
charged particles in classical non-abelian chromodynamics.   
\end{abstract}
\maketitle

\section{\label{intro} Introduction}

Physical symmetry is more often a guiding principle then a specific  
property of a system\cite{Marsden:1994, LandauC:2005}.  For example, for the 
Lagrangian of a specific system, all well established physical symmetries must be present, 
or else the resulting dynamical solutions are simply wrong. Although in some cases they may 
not be obvious, symmetries are necessarily in the equations and are typically revealed 
through conservation laws.  

The use of Lie algebras and Lie groups, permeates the theory of differential 
equations\cite{Hydon:2000}, classical and quantum mechanics\cite{Nambu:1973,Suzuki:1984}, electromagnetism\cite{Baum:1995}, fluid dynamics\cite{Bistrovic:2003}, 
magnetohydrodynamics\cite{Fuchs:1991} and high energy physics\cite{Coleman:1985}.
The purpose of this paper is to show how the standard Hamiltonian structure, with Poisson brackets 
generating a Lie algebra, describes the implications of symmetries even for those cases wherein  
the differential equations are not generated from normal friction free classical mechanical systems.  
As such these methods may be used to describe symmetries at the advanced undergraduate level without 
invoking the more abstract properties of symmetry groups. 

In Section \ref{HAM1} we introduce the Poisson bracket and its properties.  
Particular attention is given to the applications in ordinary differential equations, 
conservation laws, and the connection to Lie algebras.  Before moving to the general formalism, 
a specific example - namely spin precession - is considered.  In Section \ref{PE}, the Pauli 
Hamiltonian for a spinning charged particle in an electromagnetic field is considered and the 
dynamical equations of motion are obtained.  In Section \ref{RPM}, we carry out a similar analysis 
for a relativistic charged particle and arrive at the well known Lorentz force.  In Section \ref{NLF}, 
the relativistic charged particle is generalized to color charges and the chromodynamic Lorentz force 
is derived.

\section{Hamiltonian Structures \label{HAM1}}

Let us consider an \begin{math} n \end{math}-dimensional manifold locally described by 
generalized position coordinates
\begin{equation}
x=(x^1,\cdots,x^n),
\label{HAM1_1}
\end{equation}
and generalized momentum vectors on the cotangent planes on the manifold 
\begin{equation}
p=(p_1,\cdots,p_n).
\label{HAM1_2}
\end{equation}
In classical mechanics, the whole structure (cotangent budles over a manifold) is 
more called {\em phase space} 
\begin{equation}
(p,x)\equiv(p_1,\cdots,p_n,x^1,\cdots,x^n).
\label{HAM1_3}
\end{equation}
A real Hamiltonian function on phase space (and possibly time) of the form 
\begin{equation}
H\equiv H(p,x,t) 
\label{HAM1_4}
\end{equation}
determines the equations of motion 
\begin{equation}
\dot{x}^k =\frac{\partial H}{\partial p_k}
\ \ \ {\rm and}\ \ \ \dot{p}_k =-\frac{\partial H}{\partial x^k} 
\label{HAM1_5}
\end{equation}
For any two real functions on phase space, 
\begin{math} A(p.x) \end{math} and \begin{math} B(p.x) \end{math}, 
the Poisson bracket is defined as 
\begin{equation}
\{A,B \}\equiv \frac{\partial A}{\partial p_k}\frac{\partial B}{\partial x^k}
-\frac{\partial B}{\partial p_k}\frac{\partial A}{\partial x^k},
\label{HAM1_6}
\end{equation}
wherein the Einstein convention of summing over repeated indices is being 
employed. The following important {\em algebraic identities} hold true 
\begin{eqnarray}
\{A,B \}=-\{B,A \},
\nonumber \\ 
\{A,B+C \}=\{A,B \}+\{A,C \}, 
\nonumber \\ 
\{A,BC \}=\{A,B \}C+B\{A,C \},
\nonumber \\ 
\{A,\{B,C\}\}+\{B,\{C,A\}\}+\{C,\{A,B\}\}=0.
\label{HAM1_7}
\end{eqnarray}
Furthermore, the Poisson bracket is a derivation on both functions 
\begin{math} A(p,x) \end{math} and \begin{math} B(p,x)  \end{math} 
in that for any smooth phase functions \begin{math} f(A)  \end{math} and 
\begin{math} g(B)  \end{math} we have   
\begin{equation}
\{f(A),g(B)\}=f^\prime (A)g^\prime (B)\{A,B\}.
\label{HAM1_7der}
\end{equation}
Similarly, for partial derivatives
\begin{eqnarray}
\{f(A_1,\cdots ,A_p),g(B_1,\cdots ,B_q)\}=
\nonumber \\ 
\frac{\partial f}{\partial A_a}\ 
\frac{\partial g}{\partial B_b}\ \{A_a,B_b\}.
\label{HAM1_7d}
\end{eqnarray}

Finally, real functions \begin{math} Q\equiv Q(p,x) \end{math} 
on phase space change in time according to 
\begin{equation}
\dot{Q}=\{H,Q\}.
\label{HAM1_8}
\end{equation}
All of the above Hamiltonian structure identities are discussed in 
standard classical mechanics text books.

\subsection{Ordinary Differential Equations\label{ODE}}

It is interesting to note that an {\em arbitrary} set of ordinary differential 
equations on a manifold 
\begin{equation}
\frac{dx^k}{dt}=v^k(x^1,\cdots,x^n)\ \ \ \ \ (k=1,\cdots ,n)
\label{HAMI_9}
\end{equation}
can be written in Hamiltonian form. Employing the Hamiltonian  
\begin{equation}
H(p,x)=p_kv^k(x),
\label{HAMI_10}
\end{equation}
i.e. 
\begin{equation}
\dot{x}^k=\frac{\partial H}{\partial p_k}=v^k(x)
\label{HAMI_11}
\end{equation}
leads to Eq.(\ref{HAMI_9}) while the conjugate Hamiltonian equation 
\begin{equation}
\dot{p}_k=-\frac{\partial H}{\partial x^k}=
-\frac{\partial v^l(x)}{\partial x^k}p_l
\label{HAMI_12}
\end{equation}
is of use in the study of solution stability. The evolutionary Eq.(\ref{HAMI_9}) 
need not refer to a system wherein the Hamiltonian is directly connected 
to the energy. Dissipative irreversible equations such as those of chemical kinetics  
can also be written in Hamiltonian form. The evolutionary differential Eq.(\ref{HAMI_9}) 
in Hamiltonian form can also be applied disciplines other than physics.

\subsection{Conservation Laws \label{CL}}
If the Hamiltonian does not depend explicitly on time, 
\begin{math} H(p,x,t)\equiv H(p,x) \end{math} then the {\em energy} 
is conserved; i.e the Hamiltonian is an integral of motion,  
\begin{equation}
E=H(p,x)\ \ \ \Rightarrow \ \ \ \dot{E}=\{H,H\}=0.
\label{HAMI_13}
\end{equation}
A conserved quantity or integral of motion, \begin{math} C(p,x)  \end{math}, 
is characterized by a vanishing Poisson bracket with the Hamiltonian 
\begin{equation}
\dot{C}=\{H,C\}=0.
\label{HAMI_14}
\end{equation}  
{\bf Theorem:} The Poisson bracket of two conserved quantities 
\begin{equation}
\{C_a,C_b\}=C_{ab}
\label{HAMI_15}
\end{equation}
is also conserved; i.e. 
\begin{equation}
\{H,C_a\}=0\ \ {\rm and }\ \ \{H,C_b\}=0
\ \ \Rightarrow \ \ \{H,C_{ab}\}=0
\label{HAMI_16}
\end{equation}
{\bf Proof:} From Eqs.(\ref{HAM1_7}) it follows that 
\begin{eqnarray}
\{H,C_{ab}\}=\{H,\{C_a,C_b\}\},
\nonumber \\ 
\{H,C_{ab}\}=\{C_a,\{H,C_b\}\}-\{C_b,\{H,C_a\}\},
\nonumber \\ 
\{H,C_{ab}\}=\{C_a,0\}-\{C_b,0\}=0.
\label{HAMI_17}
\end{eqnarray}
The theorem sometimes allows for the construction of possibly new 
conserved quantities from other known conserved quantities.

\subsection{Lie Algebras}
A set of \begin{math} N \end{math} functions on phase space  
\begin{equation}
\Lambda = (\Lambda_1 ,\cdots ,\Lambda_N )
\label{HAMI_18}
\end{equation}
form a Lie algebra if the Poisson bracket of any two members of the set 
is a linear combination of members of the set; i.e. 
\begin{equation}
\{\Lambda_a,\Lambda_b\}=-\Lambda_c f^c_{ab}
\label{HAMI_19}
\end{equation}
wherein the coefficients of the linear combination 
\begin{math} f^c_{ab} \end{math} are called the {\em structure 
constants} of the algebra. For any smooth functions 
\begin{math} Q(\Lambda )  \end{math} and 
\begin{math} R(\Lambda ) \end{math} it follows from 
Eqs.(\ref{HAM1_7d}) and (\ref{HAMI_19}) that 
\begin{eqnarray}
\{Q(\Lambda ),R(\Lambda )\}=\frac{\partial Q}{\partial \Lambda_a }
\frac{\partial R}{\partial \Lambda_b }\{\Lambda_a,\Lambda_b\},
\nonumber \\ 
\{Q(\Lambda ),R(\Lambda )\}=-\Lambda_c f^c_{ab}
\frac{\partial Q}{\partial \Lambda_a }
\frac{\partial R}{\partial \Lambda_b }\ .
\label{HAMI_110}
\end{eqnarray}
The components of angular momentum provide a well known example of a Lie algebra.

\subsection{Spin \label{SP}}

As an example of a Lie algebra one may consider the spin angular momentum 
\begin{math} {\bf S}=(S_1,S_2,S_3)   \end{math}
about the center of mass of a non-relativistic rigid body rotator. The spin 
angular momentum Lie algebra is 
\begin{eqnarray}
\{S_1,S_2\}=-S_3,
\nonumber \\ 
\{S_2,S_3\}=-S_1,
\nonumber \\ 
\{S_3,S_1\}=-S_2,
\label{HAMI_111}
\end{eqnarray}
with the spin Poisson bracket version of Eq.(\ref{HAMI_110}) reading   
\begin{equation}
\{Q({\bf S}),R({\bf S})\}=-{\bf S}\cdot 
\left(\frac{\partial Q({\bf S})}{\partial {\bf S}}\times 
\frac{\partial R({\bf S})}{\partial {\bf S}}\right).
\label{HAMI_112}
\end{equation}
If the rigid body rotator Hamiltonian is described in terms of spin and 
the moment of inertia eigenvalues 
\begin{equation}
H_{\rm rigid}({\bf S})=\frac{S_1^2}{2I_1}+\frac{S_2^2}{2I_2}
+\frac{S_3^2}{2I_3},
\label{HAM_113}
\end{equation}
then the Eq.(\ref{HAM1_8}) of motion reads 
\begin{eqnarray}
{\bf \Omega }=\frac{\partial H_{\rm rigid}({\bf S})}{\partial {\bf S}},
\nonumber \\ 
\dot{\bf S}=\{H_{\rm rigid}({\bf S}),{\bf S}\},
\nonumber \\ 
\dot{\bf S}={\bf \Omega }\times {\bf S}.
\label{HAM_114}
\end{eqnarray}
Eqs.(\ref{HAM_113}) and (\ref{HAM_114}) are, of course, the Euler equations 
for the angular velocity 
\begin{math} {\bf \Omega}=(\Omega_1,\Omega_2,\Omega_3) \end{math} 
\begin{eqnarray}
I_1 \dot{\Omega }_1=(I_3-I_2)\Omega_2\Omega_3,
\nonumber \\  
I_2 \dot{\Omega }_2=(I_1-I_3)\Omega_3\Omega_1,
\nonumber \\  
I_3 \dot{\Omega }_3=(I_2-I_1)\Omega_1\Omega_2.
\label{HAM_115}
\end{eqnarray}
As an example of rigid body rotation, consider an rigid object moving through 
space in a gravitational field \begin{math} {\bf g}({\bf r}) \end{math} due to 
other massive objects. If \begin{math} \Psi ({\bf r}) \end{math} denotes the 
gravitational potential, i.e. 
\begin{math} {\bf g}({\bf r})=-{\bf grad}\Psi ({\bf r}) \end{math}, then the 
Hamiltonian has the form 
\begin{equation}
H({\bf P},{\bf R},{\bf S})=\frac{|{\bf P}|^2}{2M}
+M \Psi({\bf R})+H_{\rm rigid}({\bf S}),
\label{HAM_116}
\end{equation}
wherein \begin{math} {\bf P} \end{math}, \begin{math} {\bf R} \end{math} 
and \begin{math} {\bf S} \end{math} represent, respectively, the total momentum, 
the center of mass position and the spin. The Poisson bracket structure is thereby  
\begin{equation}
\{A,B\}=
\frac{\partial A}{\partial {\bf P}}\cdot \frac{\partial B}{\partial {\bf R}}-
\frac{\partial B}{\partial {\bf P}}\cdot \frac{\partial A}{\partial {\bf R}}-
{\bf S}\cdot \left(\frac{\partial A}{\partial {\bf S}}\times
\frac{\partial B}{\partial {\bf S}}\right). 
\label{HAM_117}
\end{equation}
The equations of motion implied by the Hamiltonian Eqs.(\ref{HAM1_8}), 
(\ref{HAM_113}), (\ref{HAM_116}), and (\ref{HAM_117}) read 
\begin{eqnarray}
\dot{\bf R}=\{H,{\bf R}\}=\frac{\bf P}{M}={\bf V}, 
\nonumber \\ 
\dot{\bf P}=\{H,{\bf P}\}=-M{\bf grad}\Psi ,
\nonumber \\ 
\ddot{\bf R}={\bf g}({\bf R}),
\nonumber \\ 
\dot{\bf S}=\{H,{\bf S}\}={\bf \Omega }\times {\bf S}.
\label{HAM_118}
\end{eqnarray}
The above rigid body space satellite example is typical of Poisson bracket structures 
which are in part canonical phase space pairs an in part a pure Lie algebra. 

\subsection{General Formalism \label{GF}}

In the {\em general case} one may write phase space as 
\begin{equation}
(p,x,\Lambda)=(p_1,\cdots ,p_n,x^1,\cdots ,x^n,\Lambda_1,\cdots \Lambda_N),
\label{HAM_119}
\end{equation}
and employ the Poisson brachet structure 
\begin{equation}
\{A,B\}=\frac{\partial A}{\partial p_k}\frac{\partial B}{\partial x^k}
-\frac{\partial B}{\partial p_k}\frac{\partial A}{\partial x^k}
-\Lambda_cf^c_{ab}\frac{\partial A}{\partial \Lambda_a}
\frac{\partial B}{\partial \Lambda_b}\ .
\label{HAM_120}
\end{equation}
For an arbitrary Hamiltonian \begin{math} H(p,x,\Lambda ,t) \end{math} 
and smooth phase space function \begin{math} Q(p,x,\Lambda ) \end{math}, 
the equation of motion is described by 
\begin{equation}
\dot{Q}=\{H,Q\}.
\label{HAM_121}
\end{equation}
This formalism allows for a classical analogue to equations of motion 
usually described only in terms quantum mechanics. 

\section{The Pauli Hamiltonian \label{PE}}

The Pauli Hamiltonian for a charged particle moving in an electromagnetic field,
\begin{equation}
{\bf E}=-\frac{1}{c}\frac{\partial {\bf A}}{\partial t}-{\bf grad}\Phi 
\ \ {\rm and}\ \ {\bf B}=curl {\bf A},
\label{PE1}
\end{equation} 
is given by \cite{LandauCTF:2001}
\begin{eqnarray}
H({\bf p},{\bf r},{\bf S})=\frac{1}{2m}
|{\bf p}-(e/c){\bf A}({\bf r},t)|^2 
\nonumber \\ 
+e\Phi ({\bf r},t)-\gamma {\bf S}\cdot {\bf B}({\bf r},t),
\ \ \ \ \ \ \ 
\label{PE2}
\end{eqnarray} 
wherein the gyromagnetic ratio \begin{math}\gamma =ge/2mc  \end{math}.
For any two phase space functions, 
\begin{math} Q({\bf p},{\bf r},{\bf S}) \end{math} and 
\begin{math} R({\bf p},{\bf r},{\bf S}) \end{math}, the generalized Poisson 
bracket is given by 
\begin{eqnarray}
\{Q,R\}=\frac{\partial Q}{\partial {\bf p}}\cdot 
\frac{\partial R}{\partial {\bf r}}-
\frac{\partial R}{\partial {\bf p}}\cdot 
\frac{\partial Q}{\partial {\bf p}}-{\bf S}\cdot 
\left(\frac{\partial Q}{\partial {\bf S}}\times 
\frac{\partial R}{\partial {\bf S}}\right). 
\label{PE3}
\end{eqnarray} 
From the Hamilton equations of motion one finds   
\begin{eqnarray}
\dot{\bf r}\equiv {\bf v}=\{H,{\bf r}\}=\frac{\partial H}{\partial {\bf p}}
=\frac{1}{m}\left({\bf p}-\frac{e}{c}{\bf A}\right),
\nonumber \\ 
\dot{\bf v}=\frac{\partial {\bf v}}{\partial t}+\{H,{\bf v}\},
\nonumber \\
\dot{\bf v}=-\frac{e}{mc}\frac{\partial {\bf A}}{\partial t}-\frac{1}{m}
{\bf grad}(e\Phi -\gamma {\bf S}\cdot {\bf B})+\frac{m}{2}\{|{\bf v}|^2,{\bf v}\},
\nonumber \\ 
\frac{m}{2}\{|{\bf v}|^2,{\bf v}\}=\frac{e}{mc}{\bf v}\times {\bf B}.\  
\label{PE4}
\end{eqnarray}
Employing Eqs.(\ref{PE1}) and (\ref{PE4}), yields the non-relativistic 
Lorentz force on a charge and an additional force due to a magnetic moment 
interacting with a magnetic field; It is 
\begin{equation}
m\dot{\bf v}=e\left({\bf E}+\frac{1}{c}{\bf v}\times {\bf B}\right)
-{\bf grad}(\gamma {\bf S}\cdot {\bf B}).
\label{PE5}
\end{equation}   
For many charged particle examples with a magnetic moment, the Lorentz force on a charge 
is sufficiently accuracte. For an uncharged particle with a magnetic moment, 
the force \begin{math} {\bf f}=-{\bf grad}(\gamma {\bf S}\cdot {\bf B}) \end{math} 
is all there is. Finally, the spin precesses according to 
\begin{math} \dot{\bf S}=\{H,{\bf S}\}  \end{math}; i.e.
\begin{equation}
\dot{\bf S}=-\gamma {\bf B}\times {\bf S}.
\label{PE6}
\end{equation}
While the Pauli Hamiltonian Eq.(\ref{PE2}) is often used in quantum mechanics, it 
virtually never used in its classical form. Quantum spins are not very easily visualized as 
classical spinning objects. Yet the classical Eqs.(\ref{PE5}) and (\ref{PE6}) of motion 
are the same as those of Heisenberg in the quantum version of the theory.

\section{Relativistic Particle Motion \label{RPM}}

We employ the proper time interval 
\begin{eqnarray}
-c^2d\tau ^2 =\eta_{\mu \nu}dx^\mu dx^\nu ,
\nonumber \\ 
\{\eta_{\mu \nu}\}={\rm diag}(+1,+1,+1,-1)\ ,
\label{RPM1}
\end{eqnarray}
wherein the space-time vector  
\begin{math} x^\mu =({\bf r},ct) \end{math} and the energy-momentum vector 
\begin{math} p_\mu =({\bf p},-E/c) \end{math} form a Lorentz invariant 
Lie algebra structure   
\begin{eqnarray}
\{A,B\}=\left(\frac{\partial A}{\partial {\bf p}}\cdot \frac{\partial B}{\partial {\bf r}}-
\frac{\partial B}{\partial {\bf p}}\cdot \frac{\partial A}{\partial {\bf r}}\right)
\nonumber \\ 
-\left(\frac{\partial A}{\partial E} \frac{\partial B}{\partial t}-
\frac{\partial B}{\partial E} \frac{\partial A}{\partial t}\right),
\nonumber \\ 
\{A,B\}=\frac{\partial A}{\partial p_\mu}\frac{\partial B}{\partial x^\mu }
-\frac{\partial B}{\partial p_\mu }\frac{\partial A}{\partial x^\mu }\ .
\label{RPM2}
\end{eqnarray}
Let us now consider the example of a charged particle.

For a point charge \begin{math} e \end{math} of mass \begin{math} m \end{math} 
moving in an electromagnetic field, 
\begin{equation}
F_{\mu \nu}=\frac{\partial A_\mu }{\partial x^\nu }-
\frac{\partial A_\nu }{\partial x^\mu }\ ,
\label{RPM3}
\end{equation}
The Lorentz invariant ``Lagrangian'' in terms of the velocity 
\begin{math} v^\mu =dx^\mu /d\tau \end{math} and space-time position 
reads \cite{LandauCTF:2001}
\begin{equation}
{\cal L}(v,x)=\frac{1}{2}m(v^\mu v_\mu - c^2)+\frac{e}{c}v^\mu A_\mu (x).
\label{RPM4}
\end{equation}
From momentum and force are derived from the Lagrangian employing 
\begin{eqnarray}
p_\mu =\frac{\partial {\cal L}}{\partial v^\mu }=mv_\mu +\frac{e}{c}A_{\mu },
\nonumber \\ 
f_\mu =\frac{\partial {\cal L}}{\partial x^\mu }=\frac{e}{c}v^\nu 
\frac{\partial A_{\nu }}{\partial x^\mu }.
\label{RPM5}
\end{eqnarray}
The Lagrangian equations of motion then read 
\begin{eqnarray}
\frac{dp_\mu }{d\tau }=f_\mu ,
\nonumber \\  
\frac{d p_\mu }{d\tau }=m\frac{d v_\mu }{d\tau }+
\frac{e}{c}\frac{\partial A_\mu }{\partial x^\nu }v^\nu . 
\label{RPM6}
\end{eqnarray}
From Eqs.(\ref{RPM3}), (\ref{RPM5}) and (\ref{RPM6}) one finds the Lorentz 
force on a charge in the Lorentz covariant form 
\begin{equation}
m\frac{d v_\mu }{d\tau }=\frac{e}{c}F_{\mu \nu }v^\nu .
\label{RPM7}
\end{equation}

One goes from the Lagrangian to the Hamiltonian viewpoint employing 
\begin{eqnarray}
{\cal H}=v^\mu \frac{\partial {\cal L}}{\partial v^\mu }-{\cal L}
=v^\mu p_\mu -{\cal L},
\nonumber \\ 
{\cal H}(p,x)=\frac{1}{2m}\left(p-\frac{e}{c}A(x)\right)^2+\frac{1}{2}mc^2.
\label{RPM8}
\end{eqnarray}
The velocity, 
\begin{equation}
v^\mu \equiv \frac{dx^\mu }{d\tau }=\{{\cal H},x^\mu \}
=\frac{1}{m}\left(p^\mu -\frac{e}{c}A^\mu (x)\right),
\label{RPM9}
\end{equation}
obeys the algebraic relations 
\begin{equation}
\{v_\mu, v_\nu \}=-\left(\frac{e}{m^2c}\right)F_{\mu \nu}
=\left(\frac{e}{m^2c}\right)F_{\nu \mu },
\label{RPM10}
\end{equation}
in accordance with Eqs.(\ref{RPM2}), (\ref{RPM3}) and (\ref{RPM9}). 
From the Lie algebra viewpoint, the Lorentz force on a charge is 
recovered via   
\begin{eqnarray}
m\frac{dv_\mu }{d\tau }=m\{{\cal H},v_\mu \},
\nonumber \\ 
m\frac{dv_\mu }{d\tau }=\frac{m^2}{2}\{v^\nu v_\nu ,v_\mu \},
\nonumber \\ 
m\frac{dv_\mu }{d\tau }=m^2\{v_\nu ,v_\mu \}v^\nu,
\nonumber \\ 
m\frac{dv_\mu }{d\tau }=\frac{e}{c}F_{\mu \nu }v^\nu .
\label{RPM11}
\end{eqnarray}
The Lorentz force on a charge equation may is employed for an abelian gauge 
field theory such as electrodynamics. A similar equation may be employed for 
nonabelian gauge field theories such as chromodynamics. 

\section{Nonabelian Lorentz Force \label{NLF}}

The Poisson bracket structure for a particle with nonabelian {\em dynamic} 
charges 
\begin{math} (e_1,\cdots ,e_N)  \end{math} is a generalization of Eq.(\ref{RPM2}), 
\begin{equation} 
\{A,B\}=\frac{\partial A}{\partial p_\mu}\frac{\partial B}{\partial x^\mu }
-\frac{\partial B}{\partial p_\mu }\frac{\partial A}{\partial x^\mu }
-c g^d_{ab}e_d\frac{\partial A}{\partial e_a}
\frac{\partial B}{\partial e_b}\ ,
\label{NLF1}
\end{equation}
as is the Hamiltonian a generalization of Eq.(\ref{RPM8}), 
\begin{equation} 
{\cal H}(p,x)=\frac{1}{2m}\left(p-\frac{e_aA^a(x)}{c}\right)^2+\frac{1}{2}mc^2,  
\label{NLF2}
\end{equation}
wherein the dynamical charges \begin{math} (e_1,\cdots ,e_N)  \end{math} 
couple into the gauge fields 
\begin{math} (A^1_\mu (x),\cdots ,A^N_\mu (x))  \end{math}. The four velocity of 
a classical particle moving through a nonabelian gauge field follows from 
\begin{equation}
v_\mu =\{{\cal H},x_\mu \}=\frac{1}{m}\left(p_\mu -\frac{e_aA^a_\mu(x)}{c}\right)
\label{NLF3}
\end{equation}

The Poisson brackets between four velocity components follows from Eqs.(\ref{NLF1}) 
and (\ref{NLF3}) according to 
\begin{eqnarray}
m^2 \{v_\mu ,v_\nu \}= 
\nonumber \\ 
-\left\{ p_\mu , \frac{e_bA^b_\nu }{c} \right\}+
\left\{p_\nu , \frac{e_aA^a_\mu}{c} \right\}
\nonumber \\ 
+\frac{1}{c^2}\{e_a,e_b\}A^a_\mu A^b_\nu ,
\nonumber \\ 
m^2 c\{v_\mu ,v_\nu \}=-e_d(\partial_\mu A^d_\nu -\partial_\mu A^d_\nu 
+ g^d_{ab}A^a_\mu A^b_\nu ).
\label{NLF4}
\end{eqnarray}
The nonabelian gauge field is thereby defined by 
\begin{equation}
F^d_{\mu \nu }=\partial_\mu A^d_\nu -\partial_\mu A^d_\nu 
+ g^d_{ab}A^a_\mu A^b_\nu ,
\label{NLF5}
\end{equation}
wherein the velocity Poison algebra may be written as 
\begin{equation}
\{v_\mu, v_\nu \}=-\left(\frac{1}{m^2c}\right)e_dF^d_{\mu \nu}
=\left(\frac{1}{m^2c}\right)e_dF^d_{\nu \mu}.
\label{NLF6}
\end{equation}
The force on a nonabelian charge follows from 
\begin{eqnarray}
m\frac{dv_\mu }{d\tau }=m\{{\cal H},v_\mu \},
\nonumber \\ 
m\frac{dv_\mu }{d\tau }=\frac{m^2}{2}\{v^\nu v_\nu ,v_\mu \},
\nonumber \\ 
m\frac{dv_\mu }{d\tau }=m^2\{v_\nu ,v_\mu \}v^\nu ;
\label{NLF7}
\end{eqnarray}
It is 
\begin{equation}
m\frac{dv^\mu}{d\tau }=\frac{1}{c}e_d F^d_{\mu \nu }v^\nu 
\label{NLF8}
\end{equation}
which is of the Lorentz form.

\section{Conclusion \label{conc}}

The classical Hamilton equations of motion have been written as a Poisson structure 
sufficiently general to handle an arbitrary sets of ordinary differential 
equations e.g. mechanical systems with forces of friction. 
The resulting simple algebraic methods were shown to be sufficiently powerful to explore 
physical symmetries employing Lie algebras.
Elementary examples included non-relativistic magnetic moment precession and the 
relativistic mechanical orbits of color 
charged particles in classical non-abelian chromodynamic fields.
The examples given above are by no means exhaustive. A closely related example is the 
relativistic spin-orbital equations of motion in a uniform electromagnetic 
field\cite{Bargmann:1959} the Lie algebraic derivation of which is left as an open 
problem for the reader.

\end{document}